\documentclass[a4paper,onecolumn,accepted=2025-03-07]{quantumarticle}
\pdfoutput=1
\usepackage[utf8]{inputenc}
\usepackage[english]{babel}
\usepackage[T1]{fontenc}
\usepackage{hyperref}

\usepackage{tikz}
\usepackage{lipsum}
\usepackage{amsmath,amssymb,bm}
\usepackage{graphicx}
\usepackage{caption}
\usepackage[subrefformat=parens]{subcaption}

\usepackage{latexsym}

\usepackage{ntheorem}
\theoremstyle{break}
\newtheorem{thm}{Theorem}[section]

\newtheorem*{thm**}{Theorem}
\newtheorem*{lem**}{Lemma}
\newtheorem*{cor**}{Corollary}
\newtheorem*{prop**}{Proposition}

\newtheorem*{defi**}{Definition}

\theoremheaderfont{\itshape}

\newtheorem*{eg**}{Example}

\newtheorem*{rmk**}{Remark}

\newcommand{\N}{\mathbb{N}}

\newcommand{\R}{\mathbb{R}}

\newcommand{\ang}[1]{\left\langle{#1}\right\rangle}

\usepackage{comment}

\newcommand{\clbr}[1]{\left \{ #1 \right \}}
\newcommand{\sqbr}[1]{\left [ #1 \right ]}
\newcommand{\ag}[1]{\left ( #1 \right )}
\newcommand{\abs}[1]{\left | #1 \right |}

\newcommand{\hid}[1]{\lambda_{#1}}
\newcommand{\pli}[3]{#1\ag{#2 | #3}}
\newcommand{\p}[1]{\pli{#1}{\lambda}{i}}
\newcommand{\st}[1]{ \clbr{\p{#1}}_{\lambda, i}}

\newcommand{\axisM}{M}
\newcommand{\axisH}{H}

\newcommand{\varK}{S}
\newcommand{\varS}{\bar{S}}
\newcommand{\varM}{M}
\newcommand{\varH}{H}
\newcommand{\triple}{{\varM, \varH,\varK}}

\newcommand{\valM}[1]{\axisM\ag{\st{#1}}}
\newcommand{\valH}[1]{\axisH\ag{\st{#1}}}
\newcommand{\beh}{s}

\newcommand{\genSt}{p^{\triple}}
\newcommand{\bh}{b_3}
\newcommand{\bmhk}{b_4}
\newcommand{\bk}{b_1}
\newcommand{\bmkb}{b_2}

\newcommand{\bmk}{b_5}

\newcommand{\pu}{u}
\newcommand{\puu}{\bar{u}}

\newcommand{\psss}{t_3}
\newcommand{\ps}{t_1}
\newcommand{\pss}{t_2}
\newcommand{\smolM}{n_0}
\newcommand{\why}{y}

\begin{document}

\title{Trade-off relations between measurement dependence and hidden information for factorizable hidden variable models}

\author{Ryo Takakura}
\email{takakura.ryo.qiqb@osaka-u.ac.jp}
\affiliation{Center for Quantum Information and Quantum Biology, Osaka University, Osaka, Japan}

\author{Kei Morisue}
\affiliation{Department of Physics, Waseda University, Tokyo, Japan}

\author{Issei Watanabe}
\affiliation{College of Systems Engineering and Science, Shibaura Institute of Technology, Saitama, Japan}

\author{Gen Kimura}
\email{gen@shibaura-it.ac.jp}
\affiliation{College of Systems Engineering and Science, Shibaura Institute of Technology, Saitama, Japan}


\begin{abstract}
The Bell theorem is explored in terms of a trade-off relation between underlying assumptions within the hidden variable model framework. 
In this paper, recognizing the incorporation of hidden variables as one of the fundamental assumptions, we propose a measure termed `hidden information' taking account of their distribution.
This measure quantifies the number of hidden variables that essentially contribute to the empirical statistics.
For factorizable models, hidden variable models that satisfy `locality' without adhering to the measurement independence criterion, we derive novel relaxed Bell-Clauser-Horne-Shimony-Holt (Bell-CHSH) inequalities. 
These inequalities elucidate trade-off relations between measurement dependence and hidden information in the CHSH scenario. 
It is also revealed that the relation gives a necessary and sufficient condition for the measures to be realized by a factorizable model.
\end{abstract}

\maketitle

\section{Introduction}\label{sec:intro}

In the history of physics, the violation of the Bell inequality \cite{Bell_inequality,PhysRevLett.23.880} is one of the most striking results.
It changed our world view completely different from the classical one: for example, we should give up local realism to explain nature.
This result, known as the Bell theorem, has been investigated from various perspectives\footnote{We refer to \cite{Bell_Aspect_2004,sep-bell-theorem,Bell2016-HEAQNA,Gisin,RevModPhys.86.419} for a comprehensive review and thorough discussion of the Bell theorem.}. 
For instance, how it is related to uncertainty relations was revealed \cite{PhysRevA.87.052125,PhysRevLett.103.230402}, and its extensions to general probabilistic theories (GPTs) \cite{Lami_PhD,Plavala_2021_GPTs,Takakura_PhD} were given \cite{1367-2630-13-6-063024,PhysRevA.89.022123,Takakura_2024}. 
In practical aspects, the existence of correlations that violate the Bell inequality was used to prove the device independent security for key distribution \cite{E91,PhysRevLett.97.120405}, and also for randomness expansion \cite{Pironio2010} and randomness extraction \cite{PhysRevA.90.032313}.


To address the Bell theorem, adopting a local hidden variable model is essential: A set of `hidden' variables is introduced that satisfies certain assumptions to explain the empirically observed statistics. 
One of the crucial assumptions behind the Bell theorem is \textit{measurement independence} \cite{Bell1985Exchange,PhysRevLett.105.250404,PhysRevLett.106.100406} (also referred to as the free-choice \cite{WisemanCavalcanti2017,doi:10.1073/pnas.1002780107} , no-superdeterminism assumptions \cite{Shimony1976,Larsson_2014}, or the existence of {\it free will} \cite{Bell1977,Brans1988}), which indicates that the choice of measurements is independent of the underlying hidden variables. 
This is often implicitly assumed, as was also the case in Bell's original paper \cite{Bell_inequality}, but it is indispensable for deriving the Bell inequality.  
Another essential assumption is \textit{locality}, which is the combination of \textit{parameter independence} and \textit{outcome independence} \cite{sep-bell-theorem,ParameterOutcomeIndependence}. 
These independences, together with measurement independence, imply that local statistics are not influenced instantaneously (or superluminally) by remote (spacelike separated) measurement settings and also remote outcomes, respectively\footnote{In the original work by Bell \cite{Bell_inequality}, {\it determinism} ({\it causality}) was also assumed. 
It is noteworthy that outcome independence holds for any deterministic hidden variable model.}.
In summary, the assumptions behind the Bell theorem can be encapsulated into the introduction of hidden variables along with three forms of independence:  (i) parameter independence, (ii) outcome independence, and (iii) measurement independence. 
In the following, we call a hidden variable model a {\it  Bell-local} model if all these assumptions are satisfied, and a {\it factorizable} model\footnote{
See Section~\ref{sec:preliminary} for the reason for adopting this terminology.
}
if only the former two assumptions (i) and (ii) are satisfied relaxing the measurement independence.
The violation of the Bell inequality thus implies that at least one of the underlying assumptions is violated. 
However, it does not tell anything about how much each assumption should be violated. 
With this background, Hall introduced quantitative measures for each assumption and derived relaxed Bell inequalities as trade-off relations between these measures \cite{PhysRevLett.105.250404,PhysRevA.82.062117,PhysRevA.84.022102}. 
More precisely, considering a deterministic hidden variable model, he introduced measures for {\it indeterminism}, {\it signaling}, and {\it measurement dependence}, and obtained a relaxed Bell-Clauser-Horne-Shimony-Holt (Bell-CHSH) inequality in the CHSH scenario. 
Since then, much research has been done in the same line of thought \cite{Banik2012,PhysRevA.99.012121,PhysRevA.104.032205,Puetz2014,PhysRevLett.109.160404,RetroCausality}. 
For instance, a relaxed Bell inequality was obtained for a given factorizable model in \cite{Puetz2014} with its upper and lower bounds given by the conditional probabilities of the measurement contexts for the underlying variables. 
In \cite{PhysRevLett.109.160404}, there was found a relaxed Bell inequality as a trade-off relation between the guessing probability and a free will parameter. 

Despite the recognition that the introduction of hidden variables is one of the fundamental assumptions behind the Bell Theorem, these studies have not considered the quantitative analysis of this assumption. 
In our previous paper \cite{Kimura_RelaxedBell}, we introduced a measure termed {\it hiddenness} to quantify the influence of introducing hidden variables, and derived a relaxed Bell inequality as a trade-off relation between the measurement dependence and the hiddenness for any factorizable model. 
However, the measure of hiddenness is limited to only discrete values and does not adequately reflect the statistics of the hidden variable model. 
In this paper, we address this issue by introducing a refined entropic measure, termed \textit{hidden information}, which takes the statistics of hidden variables into account. 
We then derive a novel relaxed Bell inequality (more precisely, a relaxed Bell-CHSH inequality as in the previous studies) for any factorizable model in the CHSH scenario (Theorem \ref{thm:main}). 
This inequality establishes a trade-off relation between measurement dependence and hidden information.
Moreover, we demonstrate that the set of inequalities that we found completely characterizes factorizable models: any point satisfying the inequality can be realized by a suitably chosen factorizable model  (Theorem \ref{thm:main2}). 
Our results give a different evaluation of the CHSH value of a factorizable model, which makes it possible in another way to interpret a violation of the usual Bell-CHSH inequality in terms of the measurement dependence and the introduction of hidden variables itself.

This paper is organized as follows.
In Section \ref{sec:preliminary}, we introduce fundamental notions on the CHSH scenario, and make a brief review on the relaxed Bell inequalities obtained by Hall \cite{PhysRevLett.105.250404,PhysRevA.82.062117,PhysRevA.84.022102}. 
In  Section \ref{sec:main}, after introducing a measure for hidden information, we show novel relaxed Bell inequalities for factorizable models, and discuss its physical interpretation.
We summarize our results in Section \ref{sec:conc}.

\section{Preliminaries}
\label{sec:preliminary}
Consider two spacelike separated parties, Alice and Bob, who perform measurements $x$ and $y$ on each subsystem to obtain outcomes $a$ and $b$ respectively. 
In the CHSH scenario, there are two choices of measurements $x, y\in\{0,1\}$ with two outcomes $a,b\in\{-1,1\}$ for each system.
The state, called a \textit{behavior} \cite{Bell_inequality}, is a set of probabilities $s=\{p(a,b|x,y)\}_{a,b\in\{-1,1\},x,y\in\{0,1\}}$, where $p(a,b|x,y)$ denotes the joint probability of observing outcomes $a$ and $b$ when measurements $x$ and $y$ are performed by Alice and Bob respectively.
We say that	a behavior $s=\{p(a,b|x,y)\}_{a,b\in\{-1,1\},x,y\in\{0,1\}}$ admits a \textit{hidden variable model} (HV model) if there exists a set $\Lambda$ of `hidden' variables through which each joint probability is given as the marginal distribution
\begin{align}
	\label{eq:hidden0}
	p(a,b|x,y)= \sum_\lambda p(a,b,\lambda|x,y),
\end{align}
where $p(a,b,\lambda|x,y)$ is a joint probability of $a,b$ and a hidden variable $\lambda \in \Lambda$ conditioned on $x,y$.
The equation \eqref{eq:hidden0} can be rewritten by using the conditional probabilities as
\begin{align}
	\label{eq:HVM}
	p(a,b|x,y)= \sum_\lambda p(\lambda|x,y) p(a,b|x,y,\lambda). 
\end{align}
Although in this paper we basically treat a discrete set of hidden variables, one can consider continuous cases as well just by replacing the summation by an integration.
In the following, we sometimes refer to the sets $I=\{\{p(\lambda|x,y)\}_{\lambda}\}_{x,y}$ and $O=\{\{p(a,b|x,y,\lambda)\}_{a,b}\}_{{x,y,\lambda}}$ as the \textit{input} and \textit{output} of the underlying HV model respectively.
We note that once two sets of probabilities $I=\{\{p(\lambda|x,y)\}_{\lambda}\}_{x,y}$ and $O=\{\{p(a,b|x,y,\lambda)\}_{a,b}\}_{{x,y,\lambda}}$ are given, an HV model $s=\{p(a,b|x,y)\}_{a,b,x,y}$ can be defined through \eqref{eq:HVM}.
Thus we write a behavior $s$ also as $s = (I,O)$ by means of its input $I$ and output $O$.

In this paper, we consider {\it factorizable} models.
An HV model $s=(I,O)$ is called factorizable if it satisfies the factorization condition
\begin{align}\label{eq:separable}
p(a,b|x,y,\lambda) = p(a|x,\lambda) p(b|y,\lambda)
\end{align}
characterized by local distributions $\{\{p(a|x,\lambda)\}_a\}_{x,\lambda}$ and $\{\{p(b|y,\lambda)\}_b\}_{y,\lambda}$.
It is worthwhile to elucidate the reasons behind this terminology.
These models were referred to as {\it measurement dependent local} in \cite{P_tz_2016} (also see \cite{Kimura_RelaxedBell}): the condition \eqref{eq:separable} is equivalent to parameter independence and outcome independence, which typically reflect the locality condition \cite{sep-bell-theorem,ParameterOutcomeIndependence}.  
However, as pointed out in \cite{RetroCausality}, relaxing measurement independence while maintaining parameter independence can cause violations of the no-signaling condition \cite{PR-box} in the empirical distributions. This observation led the authors in \cite{RetroCausality} to call these models {\it separable}, but the term `separable' is already widely recognized in the field of quantum information theory as referring quantum separable states.
To avoid confusion and more accurately reflect the characteristic that an outcome probability is factorizable under the hidden variables, in this paper we adopt the term {\it factorizable} for the models instead.

If the input $I$ of $s$ in addition satisfies the {\it measurement independence} between every measurement context $(x,y)$ and underlying variable $\lambda$, i.e., 
\begin{align}\label{eq:MI}
p(\lambda|x,y) = p(\lambda),
\end{align}
then the model is called {\it Bell-local}.
Hence the underlying assumptions for Bell-local models are three independencies: (i) parameter, (ii) outcome, and (iii) measurement  independencies of the underlying (hidden) variables. 
Bell-local models are the local HV models that have been adopted traditionally when discussing the Bell theorem. 

For a behavior $s=\{p(a,b|x,y)\}_{a,b\in\{-1,1\},x,y\in\{0,1\}}$, its CHSH value $S(s)$ is defined as 
\begin{align}
	S(s)= &\max \Bigl\{ |\ang{00}+\ang{01}+\ang{10}-\ang{11}|,|\ang{00}+\ang{01}-\ang{10}+\ang{11}|, \notag \\
	& \qquad \qquad  |\ang{00}-\ang{01}+\ang{10}+\ang{11}|,|-\ang{00}+\ang{01}+\ang{10}+\ang{11}|\Bigr\}
\end{align}
with
\begin{align}\label{eq:expectedvalue}
	\ang{xy}=\sum_{a,b\in\{-1,1\}} ab~p(a,b|x,y)\quad(x,y\in\{0,1\}).
\end{align}
Notice that the CHSH value $S(s)$ can be directly accessed by experiments.
There is a trivial upper bound 
$$
S(s) \le 4
$$
since $|\ang{xy}| \le 1$ for any $x,y\in\{0,1\}$. 
It is known that for a Bell-local behavior $s = (I,O)$ 
\begin{align}\label{eq:BCHSHineq}
S(s) = S(I,O) \le 2
\end{align}
holds, which is the famous Bell-CHSH inequality \cite{Bell_inequality,PhysRevLett.23.880}.
The violation of this inequality thus implies that at least one of the assumptions (i), (ii), or (iii) of a Bell-local model is violated. 
The most important theory that violates the Bell-CHSH inequality is quantum theory.
There, by means of a suitable choice of measurements and an entangled state, the Bell-CHSH inequality is violated \cite{PhysRevLett.23.880,PhysRevLett.103.230402} (in fact it can reach the Tsirelson bound $2 \sqrt{2}$ \cite{Tsirelson1980}), and thus quantum behaviors cannot be described by Bell-local models.

Let us focus on the CHSH value $S(s)$ of a factorizable model $s =\{p(a,b|x,y)\}_{a,b,x,y}= (I,O)$ that is not necessarily measurement independent (i.e, not necessarily Bell-local).
For the CHSH value, Hall proved a simple evaluation in terms of the measurement dependence of the model defined as \cite{PhysRevLett.105.250404,PhysRevA.84.022102}
\begin{align}
	\label{eq:meas dependency}
	M(s)=\max_{x,y,x',y'}\sum_{\lambda\in\Lambda}|p(\lambda|x,y)-p(\lambda|x',y')|.
\end{align}
It is easy to see that $0 \le M(s) \le 2$ and $M=0$ if and only if the model is measurement independent.
Introducing measures of indeterminisim and signaling as well, Hall obtained relaxed Bell inequalities (relaxed Bell-CHSH inequalities) that give trade-off relations among these measures and CHSH values \cite{PhysRevA.84.022102}.
In particular, for the factorizable model $s$, the inequality gives 
\begin{align}
	\label{eq:Hup}
	S \le 2+\min[3 M ,\ 2].
\end{align}
	Note that here and in the following $S(s)$ and $M(s)$ are denoted simply as $S$ and $M$ respectively when the behavior of interest is obvious.
The inequality \eqref{eq:Hup} can be regarded as a generalization of the Bell-CHSH inequality.
In fact, if $M = 0$ holds for a factorizable model, then the model reduces to a Bell-local model, and the Bell-CHSH inequality is reproduced as expected. 

\section{Results}
\label{sec:main}

Although a broad class of HV models has been investigated and several relaxed Bell inequalities were derived in the previous studies \cite{PhysRevLett.105.250404,PhysRevA.82.062117,PhysRevA.84.022102,Banik2012,PhysRevA.99.012121,PhysRevA.104.032205,Puetz2014,PhysRevLett.109.160404}, 
no attempt has been made to quantify the assumption of introducing hidden variables.
Motivated by this observation, in this section we introduce a measure termed {\it hidden information}, which quantifies the number of hidden variables essentially required to account for the empirical statistics.
In terms of this measure, we derive a new set of relaxed Bell inequalities for factorizable models as trade-off relations between measurement dependence and hidden information for factorizable models.
The expression of hidden information used in our argument is a refined version of the {\it hiddenness} defined originally in the previous paper \cite{Kimura_RelaxedBell}, and thus contributes to obtaining refined inequalities.
The new inequalities are also examined as trade-off relations between measurement dependence and hidden information.
We remark that in our argument we only consider the case where the number of the hidden variables is finite, i.e., the set of the hidden variables is written as $\Lambda=\{\lambda_1, \ldots, \lambda_n\}$  ($n$: finite) as well as its limiting case $ n \to \infty$.

We begin with reviewing the previous result in \cite{Kimura_RelaxedBell}.
There was defined a measure of {\it hiddenness} simply by the number (the cardinality) of hidden variables $\#(\Lambda)$. 
That is, for an HV model $s$ with a set of hidden variables $\Lambda$, its hiddenness $H'$ is defined as
\begin{align}\label{oldHid}
H'= \#(\Lambda) - 1.
\end{align}
We note that here we again follow the notation in \eqref{eq:Hup} and write $H'(s)$ simply as $H'$.
By means of this $H'$ and the measurement dependence $M$ given by \eqref{eq:meas dependency}, a refined version of \eqref{eq:Hup} for a factorizable model $s$ was derived in \cite{Kimura_RelaxedBell} as
	\begin{align}
		\label{eq:KSMbdd}
		S\le 2+\min\Bigl[ \min[H',3] M,\  2\Bigr].
	\end{align}	
We remark that this inequality gives a trade-off relation between the measurement dependence and the hiddenness: 
the less the measurement dependence $M$ is, the more the hiddenness $H'$ is, and vice versa.
In \cite{Kimura_RelaxedBell}, there was also demonstrated that any triple $(M,H',S) \in \R^3$ satisfying the inequality \eqref{eq:KSMbdd} is realized by a factorizable model.

\begin{figure}
	\includegraphics[scale=0.53]{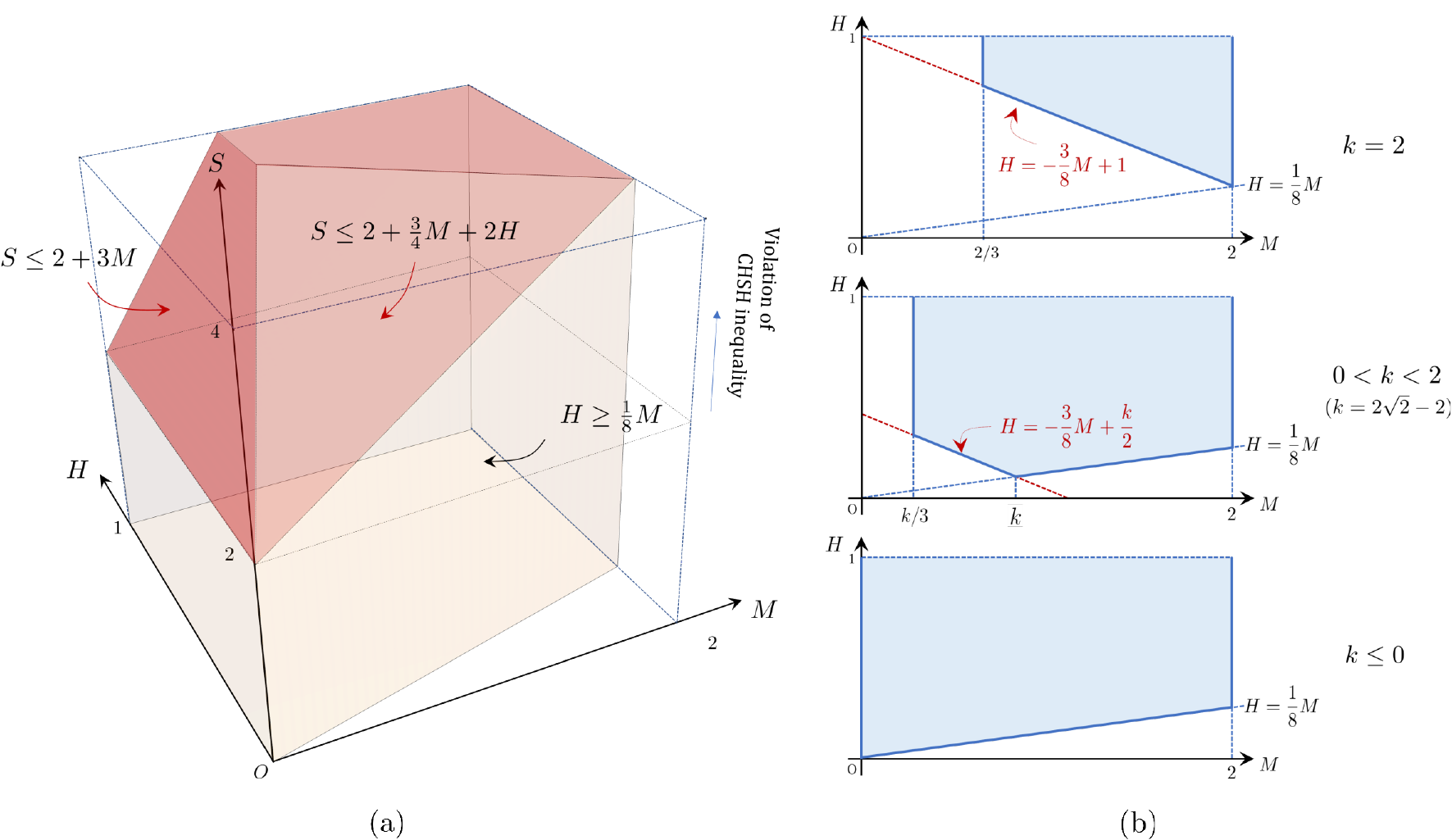}
	\caption{
		(a) The allowed region of $(M,H,S)$ in $\R^3$ determined by \eqref{eq:mains} and \eqref{eq:intrinsic}, which is a polyhedron with 10 vertices.
		The faces indicated by \eqref{eq:mains} is expressed in red and the others by \eqref{eq:intrinsic} in yellow.
		(b) The allowed regions (light blue polygone) of $(M,H)$ in $\R^2$ with $S = k + 2$ fixed: $k=2$ (the maximal violation of the Bell-CHSH inequality), $0 < k < 2$ (a violation of the Bell-CHSH inequality including the quantum case, i.e., the Tsirelson bound), and $k \le 0$ (no violation of the Bell-CHSH inequality) respectively.}
\label{fig:MHS}
\end{figure}

While the above result enables us to interpret a violation of the Bell-CHSH inequality for a factorizable model in terms of its hiddenness and measurement dependence, the measure \eqref{oldHid} would not be appropriate for quantifying hiddenness.
Two reasons can be given to show this: 
First, $H'$ is a discrete (integer-valued) quantity. 
Second, it does not always represent properly the amount of the underlying hidden variables that essentially contribute to the statistics.
To understand this, consider a factorizable model where the set of hidden variables $\Lambda$ has a large number of elements but the probability $p(\lambda)$ for each hidden variable is negligible for almost all $\lambda$'s except for small numbers of elements.
In this case, although $H'$ is inevitably estimated to be large, it is more natural to assume that there are fewer hidden variables used intrinsically.
To reflect these considerations, we introduce a new entropic measure $H$ termed {\it hidden information} by
\begin{align}
	H=1- \exp(-H_\infty(\lambda)) = 1- \max_{\lambda\in\Lambda}p(\lambda),
\end{align}
where $H_\infty(\lambda)$ is the min-entropy for the hidden variable \cite{Amigo2018}. 
Note here that the unbiased settings are concerned where measurements $x\in\{0,1\}$ and $y\in\{0,1\}$ are chosen randomly: $p(x,y)=p(x)p(y)=\frac{1}{2}\cdot\frac{1}{2}=\frac{1}{4}$, 
\begin{align}
	H=1-\frac{1}{4}\max_{\lambda\in\Lambda}\left(\sum_{x,y}p(\lambda|x,y)\right).\label{eq:hiddenness}
\end{align}
This quantity $H=1-\max_{\lambda\in\Lambda}p(\lambda)$ can be regarded as describing the `width' of the probability distribution $\{p(\lambda)\}_\lambda$ \cite{PhysRevLett.60.1103,doi:10.1063/1.3614503} and thereby it appropriately captures the necessity of hidden variables, taking the statistical distribution into account. 
It is easy to see that $H\ge0$ and $H=0$ if and only if there is an element $\lambda^* \in \Lambda$ such that $p(\lambda^*) = 1$, which essentially corresponds to the case without the introduction of hidden variables (or with a single variable).
In our finite model with arbitrary $\#(\Lambda) = n$, it holds that $H \le 1-1/n\ (<1)$, and the upper bound is attained if and only if $p(\lambda) = 1/n$ for all $\lambda\in\Lambda$. 
Note that the measurement dependence $M$ and the hidden information $H$ are functionals of the input $I=\{\{p(\lambda|x,y)\}_{\lambda}\}_{x,y}$ of the underlying HV model.
They are not completely independent of each other but intrinsically satisfy
	\begin{align}\label{eq:HM/8}
	H \ge \frac{M}{8}
	\end{align}
(the proof for this relation is given in Appendix \ref{app:MH}).
This means that large measurement dependence requires large hidden information, and also $H=0$ implies $M=0$.
Now our main finding of relaxed Bell inequalities for factorizable models is presented as follows.
\begin{thm}
\label{thm:main}
For any factorizable  model, a relaxed Bell inequality
\begin{align}\label{eq:main*}	
S \le 2+\frac{3}{4}M+2H
\end{align}
holds.
\end{thm}
This inequality is a novel trade-off relation between $M$ and $H$ for $S$ that refines the previously obtained inequality \eqref{eq:KSMbdd}. 
Combining this theorem with \eqref{eq:KSMbdd}, we obtain the tightest trade-off relation.
\begin{thm}
	\label{thm:main2}
	The measurement dependence $M$, the hidden information $H$, and the CHSH value $S$ for any factorizable model satisfy
	\begin{align}
		\label{eq:mains}		
	S \le  2+\min \Bigl[3M,\ \frac{3}{4}M+2H,\ 2\Bigr].
\end{align}
Conversely, any triple $(M,H,S) \in \R^3$ that satisfies the inequality \eqref{eq:mains} together with the intrinsic relations
\begin{align}
\label{eq:intrinsic}
0\le M\le 2,\quad \frac{M}{8} \le H < 1,\quad S\ge0
\end{align}
is realized by a factorizable model. 
\end{thm}
This theorem gives a necessary and sufficient condition for a triple $(M, H, S)$ to be realized by a factorizable model.
Fig.~\ref{fig:MHS} (a) shows the allowed region of $(M,H,S)$ for all factorizable models  determined by \eqref{eq:mains} and \eqref{eq:intrinsic}, which forms a polyhedron with $10$ vertices in $\R^3$.
The proofs of Theorems \ref{thm:main} and \ref{thm:main2} are given in Appendix \ref{app:MDF}.
	To illustrate how \eqref{eq:mains} gives a trade-off relation between $M$ and $H$, assume that a CHSH value $S=2+k$ ($-2 \le k\le2$) is observed in an experiment.
	Then, from Theorem \ref{thm:main2}, we have 
	\begin{align}
	\label{eq:bound for M}
			k\le \min \Bigl[3M,\ \frac{3}{4}M+2H,\ 2\Bigr].
	\end{align}
	It follows that the allowed region $W_{k}$ of $(M,H)$ is given by
	\begin{align}
		\label{eq:k_0 region}
		W_{k}=  \left\{
		(M, H)\mid\frac{k}{3}\le M,\ -\frac{3}{8}M+\frac{k}{2}\le H,\ 0\le M\le 2,\ \frac{M}{8} \le H<1
		\right\}.
	\end{align}
	In Fig.~\ref{fig:MHS} (b), the allowed regions $W_{k}$ for the cases $k = 2$ (the maximal violation of the Bell-CHSH inequality), $0 < k < 2$ (a violation of the Bell-CHSH inequality including the quantum case, i.e., the Tsirelson bound), and $k \le 0$ (no violation of the Bell-CHSH inequality) are plotted in light blue. 
	It implies that if the Bell-CHSH inequality is not violated ($k \le 0$), then there are no constraints between $H$ and $M$ but the intrinsic relations \eqref{eq:intrinsic}. 
    On the other hand, if the Bell-CHSH inequality is violated ($k > 0 $), then there appears a trade-off relation between $M$ and $H$ given by \eqref{eq:main*}: the less the measurement dependence $M$ is, the more the hidden information $H$ is, and vice versa.  
	In this case, we can also see that there is a lower bound for the measurement dependence $M$ given by \eqref{eq:bound for M}, which means that sufficiently large measurement dependence is needed.

\section{Conclusion}\label{sec:conc}
	In this study, we introduced a measure of hidden information where the distribution of hidden variables is taken into account. 
	We showed a novel relaxed Bell-CHSH inequality for a given factorizable model in the CHSH scenario (Theorem \ref{thm:main}), which gives a non-trivial trade-off relation between the measurement dependence and the hidden information for that model.
	The result can be considered as another version of the inequality in \cite{Kimura_RelaxedBell} in terms of another measure for hiddenness.
	Combining our novel inequality with the known trade-off relations in \cite{PhysRevA.84.022102,Kimura_RelaxedBell}, we also derived the tightest relations between measurement dependence, hidden information, and CHSH values (Theorem \ref{thm:main2}) in the sense that they give necessary and sufficient conditions for these measures to take arbitrary values realized by a factorizable model.
	Future studies will be needed to reveal how our inequalities can be re-expressed when the no-signaling condition is imposed on the empirical distribution as well as for non-factorizable hidden variable models.
	It may also be interesting to rewrite our inequalities using different measures, such as other entropic quantities.
	
	\begin{acknowledgments}
		We would like to thank M. Hall for fruitful comments and discussions. We also appreciate the anonymous referees, especially for suggesting the terminology {\it factorizable models}.
		R.T. acknowledges financial support from JSPS KAKENHI Grant No. JP21J10096, MEXT QLEAP, and JST COI-NEXT program Grant No. JPMJPF2014.
		G.K. was supported in part by JSPS KAKENHI Grant No. JP17K18107.
	\end{acknowledgments}

\appendix
\section{Proofs of Theorems \ref{thm:main} and \ref{thm:main2}}\label{app:MDF}
In this appendix, we present proofs for Theorem \ref{thm:main} in Section \ref{app:MDF1} and Theorem \ref{thm:main2} in Section \ref{app:MDF2}. 
	
\subsection{Proof of Theorem \ref{thm:main}}\label{app:MDF1}
To prove Theorem \ref{thm:main}, it is convenient to introduce the optimal CHSH value $S_{\rm opt}$ for factorizable models \cite{Kimura_RelaxedBell}  defined by
\begin{align}
	S_{\rm opt}(I)=\sup_{O\mathrm{:~ separable}} S(I, O).
\end{align}
There was also proven that for $I=\{\{p(\lambda|x,y)\}_{\lambda}\}_{x,y}$ 
\begin{align}
\label{eq:optimal bound}
S_{\rm opt}(I) = 4-2\sum_{\lambda\in\Lambda}\min_{x,y}p(\lambda|x,y)
\end{align}
holds and there exists an output $O^*$ such that $S(I, O^*)=S_{\rm opt}(I)$ (see Lemma 1 in \cite{Kimura_RelaxedBell}).
Then, because $S(s)=S(I,O)\le S_{\rm opt}(I)$ holds for any factorizable model $s=(I,O)$, our proof for Theorem \ref{thm:main} proceeds by showing that the inequality 
	\begin{align}
		\label{eq:appA}
		S_{\rm opt}(I) \le 2+\frac{3}{4}M(I)+2H(I) 
	\end{align}
	holds for any input $I=\{\{p(\lambda|x,y)\}_{\lambda\in\Lambda}\}_{x,y\in\{0,1\}}$ (note that the quantities $M$ and $H$ depend only on $I$, and thus here we write them by $M(I)$ and $H(I)$ respectively).
In the following, we represent the set $\Lambda$ of hidden variables as $\Lambda = \{\lambda_1,\lambda_2,\ldots,\lambda_n\}$ with $n = \#(\Lambda) \in \N$ (including the limiting case $n \to \infty$), and use the relabeling $i\in\{1,2,3,4\}$ for the measurement contexts $(x,y)\in\{(0,0), (0,1), (1,0), (1,1)\}$ in this order. 
	We also write the input $I=\{\{p(\lambda|x,y)\}_{\lambda\in\Lambda}\}_{x,y\in\{0,1\}}$ simply as $I=\{p(\lambda|i)\}_{\lambda\in\Lambda, i\in\{1,\ldots,4\}}$. 
	By the definitions \eqref{eq:meas dependency} and \eqref{eq:hiddenness} as well as the optimal CHSH value \eqref{eq:optimal bound}, the inequality \eqref{eq:appA} is rewritten as
	\begin{align}
		\label{eq:simple expression}
		2\tilde{K}\left(\{p(\lambda|i)\}_{\lambda, i}\right)+\frac{3}{4}\tilde{M}\left(\{p(\lambda|i)\}_{\lambda, i}\right)-\frac{1}{2}\tilde{H}\left(\{p(\lambda|i)\}_{\lambda, i}\right)\ge0,
	\end{align}
	where  
	\begin{align}
		\label{eq:KMH}
		\begin{aligned}
			&\tilde{K}\left(\{p(\lambda|i)\}_{\lambda, i}\right)=\sum_{\lambda}\min_{i}p(\lambda|i),\\
			&\tilde{M}\left(\{p(\lambda|i)\}_{\lambda, i}\right)=\max_{i,j}\left(\sum_{\lambda}|p(\lambda|i)-p(\lambda|j)|\right),\\
			&\tilde{H}\left(\{p(\lambda|i)\}_{\lambda, i}\right)=\max_{\lambda}\left(\sum_{i}p(\lambda|i)\right). 
		\end{aligned}
	\end{align}
	Thus, if we introduce
	\begin{align}\label{def:F}
		F(\{p(\lambda|i)\}_{\lambda, i}):=2\tilde{K}(\{p(\lambda|i)\}_{\lambda, i})+\frac{3}{4}\tilde{M}(\{p(\lambda|i)\}_{\lambda, i})-\frac{1}{2}\tilde{H}(\{p(\lambda|i)\}_{\lambda, i}),
	\end{align}
	then the inequality \eqref{eq:simple expression} reduces to the positivity of $F$: 
	\begin{align}
		\label{eq:simple expression0}
		F\left(\{p(\lambda|i)\}_{\lambda, i}\right)\ge0. 
	\end{align}
	Note that $\tilde{K},\tilde{M},\tilde{H}$, and hence $F$ are all invariant under any permutation of the label $i$ as well as the label $\lambda$.
	In the following, we shall prove that \eqref{eq:simple expression0} holds for any set of conditional probabilities $\{p(\lambda|i)\}_{\lambda\in\Lambda, i\in\{1,\ldots,4\}}$.

	The cases $n=1$ and $2$ are easily shown as follows. 
	When $n=1$, noting that $p(\lambda|i) =1$ for all $i$, one has $\tilde{K}=1, \tilde{M}= 0$, and $\tilde{H} = 4$.
	Thus we have 
	\begin{align*}
		F\left(\{p(\lambda|i)\}_{\lambda, i}\right)=2\cdot1-\frac{1}{2}\cdot4=0,
	\end{align*}
	which implies that \eqref{eq:simple expression0} is satisfied as an equality.
	When $n=2$, any input $\{p(\lambda|i)\}_{\lambda\in\Lambda, i\in\{1,\ldots,4\}}$ is of the form 
	\[
	\{p(\lambda|i)\}_{\lambda, i}=
	\left[
	\begin{array}{c|cccc}
		&i=1 & i=2 & i=3 &i=4\\
		\hline
		\lambda_1 & p_1& p_2 & p_3 & p_4\\
		\lambda_2 & 1-p_1& 1-p_2 & 1-p_3& 1-p_4
	\end{array}
	\right].
	\]
	By the invariance of the function $F$ under permutations of the labels $i$ and $\lambda$, we assume without loss of generality that $0 \le p_1\le p_2\le p_3\le p_4 \le 1$ and $\sum_ip_i\le4-\sum_ip_i$. 
	Direct calculations of the functions \eqref{eq:KMH} show
	\begin{align*}
		\tilde{K}(\{p(\lambda|i)\}_{\lambda, i})=p_1+(1-p_4),\ \tilde{M}(\{p(\lambda|i)\}_{\lambda, i})=2(p_4-p_1),\ \tilde{H}(\{p(\lambda|i)\}_{\lambda, i})=4-\sum_ip_i,
	\end{align*}
	and thus we obtain
	\begin{align*}
		F(\{p(\lambda|i)\}_{\lambda, i})=\frac{1}{2}(2p_1+p_2+p_3)\ge0.
	\end{align*}

	For general $n  \ge 3$, we give an inductive proof.
	Assume that \eqref{eq:simple expression0} holds for any input with $n=k(\ge2)$, and consider an input $\{p(\lambda|i)\}_{\lambda\in\{\lambda_1, \dots, \lambda_{k+1}\}, i\in\{1,\ldots,4\}}$ with $n=k+1$. 
	By the invariance of $F$ under any permutation of $\lambda$, we can assume without loss of generality that 
	\[
	\max_{\lambda\in\Lambda}\left(\sum_{i=1}^{4}p_i(\lambda)\right)=\sum_{i=1}^{4}p_i(\lambda_3),
	\]
	i.e., $\tilde{H}(\{p(\lambda|i)\}_{\lambda, i})=\sum_{i=1}^{4}p_i(\lambda_3)$ holds. 
	Let $p=\min\{p_i(\lambda)\}_{(\lambda, i)\in \Gamma}$, where $\Gamma:=\{(\lambda, i)\mid\lambda\in\{\lambda_1,\lambda_2\}, i\in\{1,2,3,4\}\}$, and assume that $p(\lambda_1|1)=p$.
	We note that the latter assumption is verified due to the invariance of $F$ under any permutation of $i$, and that it does not affect the assumption $\tilde{H}(\{p(\lambda|i)\}_{\lambda,i})=\sum_{i=1}^{4}p_i(\lambda_3)$).
	To prove the positivity of $F(\{p(\lambda|i)\}_{\lambda, i})$, let us introduce the following three transformations for the input $I=\{p(\lambda|i)\}_{\lambda,i}$. 
	The first transformation $I=\{p(\lambda|i)\}_{\lambda, i} \to I'=\{q(\lambda|i)\}_{\lambda,i}$ is given by the rule  
	\begin{align}
		\label{eq:def of new q}
		q(\lambda|i)=\left\{
		\begin{aligned}
			&p(\lambda|i)-p\quad&&(\lambda=\lambda_1)\\
			&p(\lambda|i)+p\quad&&(\lambda=\lambda_3)\\
			&p(\lambda|i)\quad&&(\mbox{otherwise}).
		\end{aligned}
		\right.
	\end{align}
	With $\Gamma':=\Gamma~\backslash\{(\lambda,i)=(\lambda_1, 1)\}$ and $\overline{\lambda_1}=\lambda_2$  and $\overline{\lambda_2}=\lambda_1$, the second transformation $I'=\{q(\lambda|i)\}_{\lambda, i} \to I''=\{r(\lambda|i)\}_{\lambda,i}$ is described by the following algorithm (A) - (C). 
	\begin{itemize}
		\item[(A)] 
		If there exists $i\in\{1,2,3,4\}$ such that $q(\lambda_1|i)=q(\lambda_2|i)=0$, then let $r(\lambda|i) = q(\lambda|i)$, and end the algorithm.
		Otherwise, go to (B).
		\item[(B)]
		Let $q:=\min(\{q(\lambda|i)\}_{(\lambda,i)\in\Gamma'})=q(\lambda^*|i^*)$ ($(\lambda^*,i^*)\in\Gamma')$), $I:=\{i\in\{1,2,3,4\}\mid q(\lambda^*|i)=0\}$, and $I^c:=\{1,2,3,4\}\backslash~I$.
		Introduce a new input $\{q'(\lambda|i)\}_{\lambda, i}$ by
		\begin{align}
			\label{eq:def of new q'}
			q'(\lambda|i)=q'(\lambda|i)=\left\{
			\begin{aligned}
				&q(\lambda|i)-q\quad&&(\mbox{$\lambda=\lambda^*, i\in I^c$})\\
				&q(\lambda|i)-q\quad&&(\mbox{$\lambda=\overline{\lambda^*}, i\in I$})\\
				&q(\lambda|i)+q\quad&&(\lambda=\lambda_3)\\
				&q(\lambda|i)\quad&&(\mbox{otherwise}),
			\end{aligned}
			\right.
		\end{align}
		and go to (C)
		(here we note that if $I=\emptyset$, then $q'(\lambda^*|i)=q(\lambda^*|i)-q$ and $q'(\overline{\lambda^*}|i)=q(\overline{\lambda^*}|i)$ for all $i$).
		\item[(C)] Let $\{q(\lambda|i)\}_{\lambda, i}=\{q'(\lambda|i)\}_{\lambda, i}$ and $\Gamma'=\Gamma'~\backslash\{(\lambda^*, i^*)\}$, and
		return to (A).
	\end{itemize}
	We can see that the algorithm is designed to make the newly obtained input $I''=\{r(\lambda|i)\}_{\lambda,i}$ satisfy $r(\lambda_1|1)=r(\lambda_2|1) = 0$.  
	The final transformation $I''=\{r(\lambda|i)\}_{\lambda, i} \to I'''=\{r'(\lambda|i)\}_{\lambda,i}$ is given by the rule
	\begin{align}
		\label{eq:def of new r'}
		r'(\lambda'_\theta|i)=\left\{
		\begin{aligned}
			&r(\lambda_1|i)+r(\lambda_2|i)\quad&&(\theta=1)\\
			&r(\lambda_{\theta+1}|i)\quad&&(\theta=2,\ldots,k).
		\end{aligned}
		\right.
	\end{align}
	Note that through the transformations,  $I'$, $I''$ and $I'''$ are all valid probabilities, which means, for example, that $\sum_{\lambda}r(\lambda|i)=1$ holds for all $i$ in $I''$.  
	Moreover, the function $F$ is monotonically decreasing through the transformations:
	\begin{align}
		F(\{p(\lambda|i)\}_{\lambda, i})\ge F(\{q(\lambda|i)\}_{\lambda, i}) \ge F(\{r(\lambda|i)\}_{\lambda, i}) \ge F(\{r'(\lambda|i)\}_{\lambda, i}).\label{eq:p ge q}
	\end{align}
	Let us show these by examining each transformation one by one.
	For the first transformation, it is easy to see that 
	\[
	\tilde{K}(\{q(\lambda|i)\}_{\lambda, i})=\tilde{K}(\{p(\lambda|i)\}_{\lambda, i}),\ 
	\tilde{M}(\{q(\lambda|i)\}_{\lambda, i})=\tilde{M}(\{p(\lambda|i)\}_{\lambda, i}),
	\]
	and
	\[
	\tilde{H}(\{q(\lambda|i)\}_{\lambda, i})=\tilde{H}(\{p(\lambda|i)\}_{\lambda, i})+4p
	\]
	hold.
	By the definition \eqref{def:F} and the non-negativity of $p$, we have $F(\{p(\lambda|i)\}_{\lambda, i})\ge F(\{q(\lambda|i)\}_{\lambda, i}) $. 
	For the second transformation, it is enough to confirm that $F(\{q'(\lambda|i)\}_{\lambda, i})\le F(\{q(\lambda|i)\}_{\lambda, i})$ holds in step (B).
	We prove this by considering the following three cases: (i) $q=0$, (ii) $q\neq0$ and $I=\emptyset$, (iii) $q\neq0$ and $I\neq\emptyset$. 
	In case (i), $\{q'(\lambda|i)\}_{\lambda, i}=\{q(\lambda|i)\}_{\lambda, i}$ holds clearly, and thus $F(\{q'(\lambda|i)\}_{\lambda, i})=F(\{q(\lambda|i)\}_{\lambda, i})$ is concluded.
	In case (ii), we can demonstrate easily that 
	\begin{align*}
		&\tilde{K}(\{q'(\lambda|i)\}_{\lambda, i})=\tilde{K}(\{q(\lambda|i)\}_{\lambda, i}),\\
		&\tilde{M}(\{q'(\lambda|i)\}_{\lambda, i})=\tilde{M}(\{q(\lambda|i)\}_{\lambda, i}),\\
		&\tilde{H}(\{q'(\lambda|i)\}_{\lambda, i})=\tilde{H}(\{q(\lambda|i)\}_{\lambda, i})+4q
	\end{align*}
	hold.
	It follows that $F(\{q'(\lambda|i)\}_{\lambda, i})=F(\{q(\lambda|i)\}_{\lambda, i})-2q\le F(\{q(\lambda|i)\}_{\lambda, i})$.
	For case (iii), we need a slightly complicated argument.
	Let us first investigate $\tilde{K}(\{q'(\lambda|i)\}_{\lambda, i})$.
	Since 
	\begin{align*}
		&\min_i q'(\lambda^*|i)=\min_i q(\lambda^*|i)=0,\\
		&\min_i q'(\overline{\lambda^*}|i)\le\min_i q(\overline{\lambda^*}|i),
		\\
		&\min_i q'(\lambda_3|i)=\min_i q(\lambda_3|i)+q,\\
		&\min_i q'(\lambda|i)=\min_i q(\lambda|i)\quad(\lambda\in\{\lambda_4, \ldots, \lambda_{k+1}\}),
	\end{align*}
	we have $\tilde{K}(\{q'(\lambda|i)\}_{\lambda, i})\le\tilde{K}(\{q(\lambda|i)\}_{\lambda, i})+q$.
	We next focus on $\tilde{M}(\{q'(\lambda|i)\}_{\lambda, i})$.
	The following observations are important (remember that $q(\lambda^*|i)=0$ for $i\in I$):
	\begin{align*}
		&|q'(\lambda^*|i)-q'(\lambda^*|j)|
		=
		\left\{
		\begin{aligned}
			&|q(\lambda^*|i)-q(\lambda^*|j)|\quad(\mbox{$i,j\in I$ or $i,j\in I^c$})\\
			&|q(\lambda^*|i)-q(\lambda^*|j)|-q\quad(\mbox{$i\in I$ and $j\in I^c$}),
		\end{aligned}
		\right.\\
		&|q'(\overline{\lambda^*}|i)-q'(\overline{\lambda^*}|j)|
		=
		\left\{
		\begin{aligned}
			&|q(\overline{\lambda^*}|i)-q(\overline{\lambda^*}|j)|\quad(\mbox{$i,j\in I$ or $i,j\in I^c$})\\
			&|q(\overline{\lambda^*}|i)-q(\overline{\lambda^*}|j)\pm q|\quad(\mbox{$i\in I$ and $j\in I^c$}),
		\end{aligned}
		\right.\\
		&|q'(\lambda|i)-q'(\lambda|j)|
		=|q(\lambda|i)-q(\lambda|j)|\quad(\mbox{for all $i,j$ and $\lambda\in\{\lambda_3, \ldots, \lambda_{k+1}\}$}).
	\end{align*}
	Because $|q(\overline{\lambda^*}|i)-q(\overline{\lambda^*}|j)\pm q|\le|q(\overline{\lambda^*}|i)-q(\overline{\lambda^*}|j)|+q$, we can see from these observations that 
	\[
	\sum_{\lambda\in\Lambda}|q'(\lambda|i)-q'(\lambda|j)|\le\sum_{\lambda\in\Lambda}|q(\lambda|i)-q(\lambda|j)|
	\] 
	holds for all $i,j$, which implies $\tilde{M}(\{q'(\lambda|i)\}_{\lambda, i})\le\tilde{M}(\{q(\lambda|i)\}_{\lambda, i})$.
	For $\tilde{H}(\{q'(\lambda|i)\}_{\lambda, i})$, we obtain easily $\tilde{H}(\{q'(\lambda|i)\}_{\lambda, i})=\tilde{H}(\{q'(\lambda|i)\}_{\lambda, i})+4q$.
	Therefore, we can conclude that 
	$F(\{q'(\lambda|i)\}_{\lambda, i})\le F(\{q(\lambda|i)\}_{\lambda, i})$
	holds also in case (iii). 
	The third inequality in \eqref{eq:p ge q} as well as the positivity of $F(\{r'(\lambda|i)\}_{\lambda, i})$ is shown as follows. 
	First, it is easily verified 
	\begin{align*}
		&\sum_{\lambda'\in\Lambda'} \min_i r'(\lambda'|i)=\sum_{\theta=3}^{k+1}\min_i r(\lambda_\theta|i)=\sum_{\lambda\in\Lambda} \min_i r(\lambda|i),\\
		&\max_{\lambda'\in\Lambda'} \left(\sum_i r'(\lambda'|i)\right)\ge \max_{\lambda\in\Lambda} \left(\sum_i r(\lambda|i)\right),
	\end{align*}
	which implies
	\begin{align*}
		&\tilde{K}(\{r'(\lambda'|i)\}_{\lambda', i})=\tilde{K}(\{r(\lambda|i)\}_{\lambda, i}),\\
		&\tilde{H}(\{r'(\lambda'|i)\}_{\lambda', i})\ge\tilde{H}(\{r(\lambda|i)\}_{\lambda, i}).
	\end{align*}
	In addition, because
	\begin{align*}
		|r'(\lambda'_\theta|i)-r'(\lambda'_\theta|j)|=\left\{
		\begin{aligned}
			&|(r(\lambda_1|i)+r(\lambda_2|i))-(r(\lambda_1|j)+r(\lambda_2|j))|\quad&&(\theta=1)\\
			&|r(\lambda_{\theta+1}|i)-r(\lambda_{\theta+1}|j)|\quad&&(\theta=2,\ldots,k),
		\end{aligned}
		\right.
	\end{align*}
	and 
	\begin{align*}
		|(r(\lambda_1|i)+r(\lambda_2|i))-(r(\lambda_1|j)+r(\lambda_2|j))|\qquad\qquad\qquad\\
		\le|r(\lambda_1|i)-r(\lambda_1|j)|+|r(\lambda_2|i)-r(\lambda_2|j)|,
	\end{align*}
	we have 
	\[
	\tilde{M}(\{r'(\lambda'|i)\}_{\lambda', i})\le\tilde{M}(\{r(\lambda|i)\}_{\lambda, i}),
	\]
	and thus $F(\{r'(\lambda'|i)\}_{\lambda', i})\le F(\{r(\lambda|i)\}_{\lambda, i})$. 
	Together with the assumption $F(\{r'(\lambda'|i)\}_{\lambda', i})\ge0$ for the table $\{r'(\lambda'|i)\}_{\lambda', i}$ with $k$ rows, this inequality implies $F(\{r(\lambda|i)\}_{\lambda, i})\ge0$.
	Since $F(\{r(\lambda|i)\}_{\lambda, i})\le F(\{q(\lambda|i)\}_{\lambda, i})$ and $F(\{q(\lambda|i)\}_{\lambda, i})\le F(\{p(\lambda|i)\}_{\lambda, i})$, we can conclude $F(\{p(\lambda|i)\}_{\lambda, i})\ge0$. 
	
	Finally, we prove that the claim $F(\{p(\lambda|i)\}_{\lambda, i})\ge0$ holds also for $\{p(\lambda|i)\}_{\lambda\in\Lambda, i=1,\ldots,4}$ with countably infinite $\Lambda=\{\lambda_1,\lambda_2,\ldots\}$
	(note that in this case the definition of $H$ is modified as $H=1-\sup_{\lambda\in\Lambda}p(\lambda)$).
	To see this, we first confirm that all the quantities in \eqref{eq:KMH} have definite values for this $\{p(\lambda|i)\}_{\lambda\in\Lambda, i=1,\ldots,4}$.
	In fact, for example, we can see that the right hand side of
	\[
	\tilde{K}\left(\{p(\lambda|i)\}_{\lambda, i}\right)=\sum_{\lambda\in\Lambda}\min_{i}p(\lambda|i)
	\]
	is definite because $\sum_{\lambda\in\Lambda}\min_{i}p(\lambda|i)\le\sum_{\lambda\in\Lambda}p(\lambda|i)=1$.
	Now we construct a family of tables $\{\{p_\alpha(\lambda|i)\}_{\lambda\in\Lambda, i=1,\ldots,4}\}_{\alpha=1, 2,\ldots}$ by 
	\begin{align*}
		p_\alpha(\lambda|i)=\left\{
		\begin{aligned}
			&p(\lambda_\alpha|i)\quad&&(\lambda=\lambda_1, \ldots, \lambda_{\alpha-1})\\
			&1-\sum_{\beta=1}^{\alpha-1}p_\alpha(\lambda_{\beta}|i)\quad&&(\lambda=\lambda_\alpha)\\
			&0\quad&&(\lambda=\lambda_{\alpha+1}, \lambda_{\alpha+2}, \ldots).
		\end{aligned}
		\right.
	\end{align*}
	We note that each $\{p_\alpha(\lambda|i)\}_{\lambda, i}$ is essentially a finite table, and thus $F(\{p_\alpha(\lambda|i)\}_{\lambda, i})\ge0$.
	Then, because
	\begin{align*}
		\tilde{K}\left(\{p(\lambda|i)\}_{\lambda, i}\right)
		&=\sum_{\lambda\in\Lambda}\min_{i}p(\lambda|i)\\
		&=\lim_{\alpha\to\infty}\sum_{\lambda\in\Lambda}\min_{i}p_\alpha(\lambda|i)\\
		&=\lim_{\alpha\to\infty} \tilde{K}\left(\{p_\alpha(\lambda|i)\}_{\lambda, i}\right)
	\end{align*}
	holds for example, we have $\lim_{\alpha\to\infty}F(\{p_\alpha(\lambda|i)\}_{\lambda, i})=F(\{p(\lambda|i)\}_{\lambda, i})$, and $F(\{p(\lambda|i)\}_{\lambda, i})\ge0$ is concluded.

\subsection{Proof of Theorem \ref{thm:main2}}
	\label{app:MDF2}
Since the inequality \eqref{eq:mains} is derived easily from \eqref{eq:KSMbdd} and \eqref{eq:main*}, we prove the existence of a factorizable model $s$ such that
output $O^{\triple}$
	\begin{equation}\label{eq:MHS}
	  \axisM\ag{\beh}=\varM,\  \axisH\ag{\beh}=\varH,\  S\ag{\beh}=\varK
	\end{equation}
holds for any triple $(M, H, S)\in\R^3$ that satisfies the inequality \eqref{eq:mains} and \eqref{eq:intrinsic}, i.e., 
	\begin{align}
		\label{ineq5}
		\begin{aligned}
			0\leq \bk (\varK):=4-\varK, \ \
			0\leq \bmkb(\varM,\varK):=3\varM+2-\varK,\ \ \\
			0 < \bh (\varH):=1-\varH,\ 
			0\leq \bmhk (\varM,\varH, \varK):=\varH -\frac{1}{2}\varK+ \frac{3}{8}\varM+1,\\
			0\leq \varK, \ \  0\leq \axisM, \ \ \ \ 0\leq 2-\axisM, \ \ 0\leq \axisH-\axisM/8
		\end{aligned}
	\end{align}
(see Fig.~\ref{fig:MHS} (a)).

	\subsubsection{The notations of values}
	In this part, we introduce several values that are needed to prove the claim.
	Let $(M, H, S)\in\R^3$ be a triple that satisfies \eqref{ineq5}.
	We first introduce a slightly larger CHSH value $\varS\geq \varK$ by
	\begin{eqnarray}
	\label{barS}
	    \varS:=\varK + \min \clbr{\bk (\varK), \bmkb(\varM,\varK), 2\bmhk (\varM,\varH, \varK)}
	\end{eqnarray}
	so that the inequalities 
	\begin{eqnarray*}\label{ineq5alt}
	0\leq \bk (\varS),\ \  0\leq \bmkb(\varM,\varS),\ \  0\leq \bmhk (\varM,\varH, \varS)  
	\end{eqnarray*}
	in \eqref{ineq5} that are concerning to $\varS$ are satisfied still and in addition at least one of the equalities holds among them.
	Then, since another auxiliary inequality
	\begin{equation*}
	    0\leq\bmk\ag{\varM,\varS}:=\varS -\varM-2
	\end{equation*}
	holds due to \eqref{ineq5} and \eqref{barS} and natural numbers $n, n_0\in \N$ such that 
	\begin{align*}
			n \geq \frac{1}{1-\varH}\ \ \ag{\Leftrightarrow \bh\ag{\varH}-\frac{1}{n}\geq 0},\ \ 
			n =4\smolM
	\end{align*}
	always exist, we can introduce successfully non-negative numbers 
	\begin{equation*}
	    \pu:=\frac{\bh(\varH)-\frac{1}{n}}{\bmhk(\varM, \varH, \varS)+\bh(\varH)-\frac{1}{n}},\ \ 
			\puu:=\frac{\bmhk(\varM,\varH, \varS)}{\bmhk(\varM, \varH,\varS)+\bh(\varH)-\frac{1}{n}},
	\end{equation*}
	\begin{equation*}
			\ps:=\frac{\bk(\varS)}{2},\ \ 
			\pss :=\frac{\bmkb(\varM,\varS)}{4},\ \ 
			\psss :=\frac{3\bmk(\varM,\varS)}{4},
	\end{equation*}
	\begin{equation*}
	    	\why := 1/(12\smolM),
	\end{equation*}
	which satisfy $\pu+\puu=1,\ \ps+\pss+\psss=1,\  12\smolM y=1$.

	\subsubsection{The input}
	By means of the values above, let us define an input $I^{\triple} = \{\p{\genSt}\}_{\lambda,i}$ by
	\begin{eqnarray}
	\label{p^MHS}
	\begin{aligned}
		\pli{\genSt}{\hid{l}}{1} &:=\left\{
		\begin{array}{c c}
			\pu\ps +3\ps\puu\why & (l=1) \\
			3\ps\puu\why & (1<l\leq\smolM) \\ 
			3\puu\ps\why + 12\pss\why + 4\psss\why & (\smolM<l\leq 2\smolM) \\
			3\puu\ps\why  + 4\psss\why & (2\smolM<l\leq 3\smolM)\\
			3\puu\ps\why  + 4\psss\why & (3\smolM<l\leq 4\smolM)
		\end{array}
		\right.,\\
		\pli{\genSt}{\hid{l}}{2} &:=\left\{
		\begin{array}{c c}
			\pu\ag{1-\frac{2\psss}{3}} +\puu\why\ag{3\ps+12\pss+4\psss} & (l=1) \\
			\puu\why\ag{3\ps+12\pss+4\psss} & (1<l\leq\smolM) \\ 
			3\ps\puu\why & (\smolM<l\leq 2\smolM) \\
			3\puu\ps\why  + 4\psss\why & (2\smolM<l\leq 3\smolM)\\
			3\puu\ps\why  + 4\psss\why & (3\smolM<l\leq 4\smolM)
		\end{array}
		\right.,\\
		\pli{\genSt}{\hid{l}}{3} &:=\left\{
		\begin{array}{c c}
			\pu\ag{1-\frac{2\psss}{3}} +\puu\why\ag{3\ps+4\psss} & (l=1) \\
			\puu\why\ag{3\ps+4\psss} & (1<l\leq\smolM) \\ 
			3\puu\ps\why  + 4\psss\why & (\smolM<l\leq 2\smolM) \\
			3\puu\ps\why & (2\smolM<l\leq 3\smolM)\\
			4\psss\why  + \puu\why\ag{3\ps+12\pss } & (3\smolM<l\leq 4\smolM)
		\end{array}
		\right.,\\
		\pli{\genSt}{\hid{l}}{4} &:=\left\{
		\begin{array}{c c}
			\pu\ag{1-\frac{2\psss}{3}} +\puu\why\ag{3\ps+4\psss} & (l=1) \\
			\puu\why\ag{3\ps+4\psss} & (1<l\leq\smolM) \\ 
			3\puu\ps\why  + 4\psss\why & (\smolM<l\leq 2\smolM) \\
			4\psss\why  + \puu\why\ag{3\ps+12\pss } & (2\smolM<l\leq 3\smolM)\\
			3\puu\ps\why & (3\smolM<l\leq 4\smolM)
		\end{array}
		\right.	.
		\end{aligned}
	\end{eqnarray}
Note that due to \eqref{ineq5} and \eqref{ineq5alt}, this input is definitely a set of probability distributions, that is, all components are non-negative and for each $i=1,\ldots,4$, they sum up to $1$.
Now we can prove that a behavior $s$ whose input is given by $I^{\triple}$ realizes $M(s)=M$ and $H(s)=H$ in \eqref{eq:MHS}.
In fact, since $M\ag{\beh}$ and $H\ag{\beh}$ are independent of the output of $s$ and 
 \begin{align*}
	\max_{i,j}\sum_{l=1}^{n}\abs{\pli{\genSt}{\hid{l}}{i}-\pli{\genSt}{\hid{l}}{j}}&=\sum_{l=1}^{n}\abs{\pli{\genSt}{\hid{l}}{1}-\pli{\genSt}{\hid{l}}{2}},\\
	\max_{l=1,2,\dots n}\sum_{i=1}^{4}\pli{\genSt}{\hid{l}}{i}&=\sum_{i=1}^{4}\pli{\genSt}{\hid{1}}{i}
\end{align*}
hold (see \eqref{p^MHS}), direct calculations show
	\begin{align*}
		\axisM\ag{{\beh}}&= \max_{i,j}\sum_{l=1}^{n}\abs{\pli{\genSt}{\hid{l}}{i}-\pli{\genSt}{\hid{l}}{j}}\\
        &=\sum_{l}\abs{\pli{\genSt}{\hid{l}}{1}-\pli{\genSt}{\hid{l}}{2}}\\
		&=\sqbr{\pu\pss+\frac{\pu\psss}{3} + 12\why \puu\pss+4\why\puu\psss}+\ag{\smolM-1}\sqbr{12\why \puu\pss+4\why\puu\psss}\nonumber\\
		&\qquad+\smolM\sqbr{12\pss\why + 4\psss\why}+\smolM\sqbr{0}+\smolM\sqbr{0}\\
		&=\ag{\pss+\frac{\psss}{3}}\ag{1+\pu+\puu}\\
		&=\frac{\varM}{2}\cdot 2=\varM,
	\end{align*}
and
	\begin{align*}
		\axisH{\ag{\beh}}&= 1-\frac{1}{4}\max_{l=1,\dots n}\sum_{i=1}^{4}\pli{\genSt}{\hid{l}}{i}\\
        &=1-\frac{1}{4}\sum_{i=1}^{4}\pli{\genSt}{\hid{1}}{i}\\
		&=1-\frac{1}{4}\left[
		\pu\ps +3\ps\puu\why
		+\pu\ag{1-\frac{2\psss}{3}} +\puu\why\ag{3\ps+12\pss+4\psss} \right.\nonumber\\
		&\qquad\left.+\pu\ag{1-\frac{2\psss}{3}} +\puu\why\ag{3\ps+4\psss}
		+\pu\ag{1-\frac{2\psss}{3}} +\puu\why\ag{3\ps+4\psss}\right]\\
		&=1-\frac{\pu\ag{\bmhk+\bh}+\puu/\smolM}{4}\\
		&=1-\bh =\varH.
	\end{align*}

	\subsubsection{The output}
	In the argument above, we showed that any behavior $s$ whose input is $I^{\triple} = \{\p{\genSt}\}_{\lambda,i}$ defined in \eqref{p^MHS} gives $M(s)=M$ and $H(s)=H$.
	In this part, we prove the existence of a factorizable output $O^{M,H,S}$ such that the behavior $s=(I^{\triple}, O^{\triple})$ realizes the remaining condition $S(s)=S$.
	We first consider auxiliary outputs 
	\begin{align*}
	\underline{O}^{\triple}&:=\ag{\clbr{\underline{p}^{\triple}\ag{a|x,\lambda}}_{a}, \clbr{\underline{p}^{\triple}\ag{b|y,\lambda}}_{b}},\\
	\overline{O}^{\triple}&:=\ag{\clbr{\bar{p}^{\triple}\ag{a|x,\lambda}}_{a}, \clbr{\bar{p}^{\triple}\ag{b|y,\lambda}}_{b}},
	\end{align*}
	which are designed to have the CHSH values
	\begin{align}
		\label{O^bar}
		\begin{aligned}
					\varK\ag{I^\triple, \underline{O}^\triple}&=0,\\
			\varK\ag{I^\triple, \overline{O}^\triple}&=\bar{S}
		\end{aligned}
	\end{align}
respectively.
	The first auxiliary output $\underline{O}^\triple$ is constructed via
	\begin{align*}
	\underline{p}\ag{a|x,\lambda}=1/2\ \ ,
	\underline{p}\ag{b|y,\lambda}=1/2,\\
	\forall a,b=\pm 1, \forall x\in\clbr{0,1}, \forall \lambda=\lambda_1,\dots. \lambda_n,
	\end{align*}
	because it lets all the expectation values in \eqref{eq:expectedvalue} vanish.
	On the other hand, since we have
    \begin{eqnarray*}
        \min_{i}{p^{M,H,S}(\lambda_l|i)}=
        \left\{
        \begin{array}{cc}
             p^{M,H,S}(\lambda_l|1)&  \ag{1\leq l \leq \smolM}\\
             p^{M,H,S}(\lambda_l|2)& \ag{\smolM< l \leq 2\smolM}\\
             p^{M,H,S}(\lambda_l|3)&  \ag{2\smolM < l \leq 3\smolM}\\
             p^{M,H,S}(\lambda_l|4)& \ag{3\smolM < l \leq 4\smolM}\\
        \end{array}
        \right.
    \end{eqnarray*}
and thus
    \begin{align*}
        S_{\mathrm{opt}}\ag{I^\triple}
        &=4-2\sum_{l=1}^{n}\min_{i}\ag{p^{M,H,S}(\lambda_l|i)}\\
        &=4-2\sum_{l=1}^{n_0}\sum_{i=1}^{4}\pli{\genSt}{\hid{i(\smolM-1) +l}}{i}\\
        &=4-2\sqbr{\pu\ps +3\ps\puu\why+\ag{\smolM-1}\ag{3\ps\puu\why}}-2\smolM\sqbr{3\cdot 3\ps\puu\why}\\
        &=4-2\ps=4-2\cdot\frac{4-\varS}{2}=\varS
    \end{align*}
by virtue of \eqref{eq:optimal bound}, the existence of $\overline{O}^\triple=\ag{\clbr{\bar{p}^{\triple}\ag{a|x,\lambda}}_{a}, \clbr{\bar{p}^{\triple}\ag{b|y,\lambda}}_{b}}$ satisfying the second equation of \eqref{O^bar} is verified according to the previous result in \cite{Kimura_RelaxedBell} (see the description after \eqref{eq:optimal bound}).
Now let us consider a family $\{O^{\triple}_{t_A, t_B}\}_{(t_A, t_B)\in[0,1]^2}$ of factorizable outputs defined by
	\begin{eqnarray*}
	    O^{\triple}_{t_A, t_B}:=\ag{\clbr{p^{\triple}_{t_A}\ag{a|x,\lambda}}_{a}, \clbr{p^{\triple}_{t_B}\ag{b|y,\lambda}}_{b}},
	\end{eqnarray*}
	where the two local distributions are convex combinations of the corresponding two auxiliary outputs: 
	\begin{align*}
	    p^{\triple}_{t_A}\ag{a|x,\lambda}&:=\ag{1-t_A}\underline{p}\ag{a|x,\lambda}+t_A\bar{p}^{\triple}\ag{a|x,\lambda},\\
	    p^{\triple}_{t_B}\ag{b|y,\lambda}&:=\ag{1-t_B}\underline{p}\ag{b|y,\lambda}+t_B\bar{p}^{\triple}\ag{b|y,\lambda}.
	\end{align*}
Based on this family of outputs, we can introduce a function $S\ag{\ag{I^\triple, O^\triple_{t_A, t_B}}}=:S(t_A, t_B)$, which is continuous at any $(t_A,t_B) \in \sqbr{0,1}^2$ and satisfies $0(=S(0,0))\le S(t_A, t_B)\le\varS(=S(1, 1))$.
Then the intermediate-value theorem \cite{iwanami_vol1}
 guarantees that there exists $(t_A^*, t_B^*)\in \sqbr{0,1}^2$ for every $S\in[0, \varS]$ such that $S(t_A^*, t_B^*)=S$, and thus, by letting $O^{M,H,S}=O^{\triple}_{t_A^*, t_B^*}$, we obtain a factorizable behavior $s=s\ag{I^\triple, O^\triple}$ that realizes all of the conditions \eqref{eq:MHS}.

	\section{Proof of \eqref{eq:HM/8}}\label{app:MH}
	In this appendix, under the same notation as Appendix \ref{app:MDF}, we prove that 
	\begin{align}
		\label{eq:appB}
		\valH{p}\ge\frac{1}{8}\valM{p}
	\end{align}
	holds for any input $\{p(\lambda|i)\}_{\lambda,i}$.
	If $n:=\#(\Lambda)=1$, then, as we have seen in Appendix \ref{app:MDF}, $H=M=0$ holds, and thus \eqref{eq:appB} is verified.
	The proof for $n\ge2$ is again given by induction for the number of the hidden variables $n$.
	We note that the claim \eqref{eq:appB} can be rewritten explicitly as
	\begin{align}
		\label{eq:appB2}
		\frac{1}{4}\max_{\lambda}\sum_ip(\lambda|i)+\frac{1}{8}\max_{i,j}\sum_{\lambda}|p(\lambda|i)-p(\lambda|j)|\le 1,
	\end{align}
	and it is easy to see that the quantities $\max_{\lambda}\sum_ip(\lambda|i)$ and $\max_{i,j}\sum_{\lambda}|p(\lambda|i)-p(\lambda|j)|$ are invariant under permutations of the labels $i$ and $\lambda$.
	We first examine the case $n=2$.
	Any input $I=\{\{p(\lambda|i)\}_{\lambda=\lambda_1,\lambda_2, i=1,2,3,4}$ is expressed as 
	\[
	\{p(\lambda|i)\}_{\lambda, i}=
	\left[
	\begin{array}{c|cccc}
		&i=1 & i=2 & i=3 &i=4\\
		\hline
		\lambda_1 & p_1& p_2 & p_3 & p_4\\
		\lambda_2 & 1-p_1& 1-p_2 & 1-p_3& 1-p_4
	\end{array}
	\right].
	\]
	In this expression, we can assume $p_1\ge p_2$, $\sum_ip_i\le4-\sum_ip_i$, and $\max_{i,j}\sum_{\lambda}|p(\lambda|i)-p(\lambda|j)|=\sum_{\lambda}|p(\lambda|1)-p(\lambda|2)|$ without loss of generality.
	Then, since
	\[
	\sum_{\lambda}|p(\lambda|1)-p(\lambda|2)|=|p_1-p_2|+|p_1-p_2|=2(p_1-p_2)
	\]
	holds, we have 
	\begin{align*}
		\frac{1}{4}\max_{\lambda}\sum_ip(\lambda|i)+\frac{1}{8}\max_{i,j}\sum_{\lambda}|p(\lambda|i)-p(\lambda|j)|
		&=
		\left(1-\frac{1}{4}\sum_i p_i\right)+\frac{1}{4}(p_1-p_2)\\
		&=1-\frac{1}{4}(2p_2+p_3+p_4)\le 1.
	\end{align*}
	We next suppose that \eqref{eq:appB2} holds for any inputs with $n=k\ (\ge2)$, and consider an input $\{p(\lambda|i)\}_{\lambda,i}$ with $n=k+1$.
	By means of suitable permutations, this input $\{p(\lambda|i)\}_{\lambda,i}$ can be assumed to satisfy $\max_{i,j}\sum_{\lambda\in\Lambda}|p(\lambda|i)-p(\lambda|j)|=\sum_{\lambda}|p(\lambda|1)-p(\lambda|2)|$, and $p(\lambda_1|1)\ge p(\lambda_1|2)$ and $p(\lambda_2|1)\ge p(\lambda_2|2)$.
	We note that the latter condition is valid because $n=k+1\ge3$.
	Now we define an input $\{p'(\lambda'|i)\}_{\lambda\in\Lambda', i=1,\ldots,4}$ with $\Lambda'=\{\lambda'_1, \ldots, \lambda'_k\}$ ($n=k$) by
	\begin{align*}
		p'(\lambda'_\theta|i)=\left\{
		\begin{aligned}
			&p(\lambda_1|i)+p(\lambda_2|i)\quad&&(\theta=1)\\
			&p(\lambda_{\theta+1}|i)\quad&&(\theta=2,\ldots,k).
		\end{aligned}
		\right.
	\end{align*}
	It follows easily from this definition of $\{p'(\lambda'_\theta|i)\}_{\lambda',i}$ that 
	\begin{align}
		\label{eq:appB H}
		\max_{\lambda'\in\Lambda'}\sum_ip(\lambda'|i)\ge\max_{\lambda\in\Lambda}\sum_ip(\lambda|i).
	\end{align}
	To evaluate the second term of the l.h.s. of \eqref{eq:appB2}, 
	it should be noted that 
	\begin{align*}
		\sum_{\lambda'\in\Lambda'}|p(\lambda'|i)-p(\lambda'|j)|
		&=|p(\lambda'_1|i)-p(\lambda'_1|j)|+\sum_{\lambda'\in\Lambda'\backslash\{\lambda'_1\}}|p(\lambda'|i)-p(\lambda'|j)|\\
		&\le|p(\lambda_1|1)-p(\lambda_1|2)|+|p(\lambda_2|1)-p_i(\lambda_2|2)|+\sum_{\lambda\in\Lambda\backslash\{\lambda_1, \lambda_2\}}|p(\lambda|i)-p(\lambda|j)|\\
		&=
		\sum_{\lambda\in\Lambda}|p(\lambda|i)-p(\lambda|j)|\\
		&\le
		\sum_{\lambda\in\Lambda}|p(\lambda|1)-p(\lambda|2)|
	\end{align*}
	holds for any $(i,j)$, where the first inequality follows from the triangle inequality and the second one from the assumption imposed on $\{p(\lambda_\theta|i)\}_{\lambda,i}$.
	While this implies
	\begin{align*}
		\max_{i,j}\sum_{\lambda'\in\Lambda'}|p(\lambda'|i)-p(\lambda'|j)|\le \sum_{\lambda\in\Lambda}|p(\lambda|1)-p(\lambda|2)|,
	\end{align*}
	we can show
	\[
	\sum_{\lambda'\in\Lambda'}|p(\lambda'|1)-p(\lambda'|2)|= \sum_{\lambda\in\Lambda}|p(\lambda|1)-p(\lambda|2)|, 
	\]
	i.e., 
	\begin{align}
		\label{eq:appB M}
		\max_{i,j}\sum_{\lambda'\in\Lambda'}|p(\lambda'|i)-p(\lambda'|j)|=\max_{i,j}\sum_{\lambda\in\Lambda}|p(\lambda|i)-p(\lambda|j)|.	
	\end{align}
	In fact, because
	\begin{align*}
		|p(\lambda'_1|1)-p(\lambda'_1|2)|
		&=|(p(\lambda_1|1)+p(\lambda_2|1))-(p(\lambda_1|2)+p_i(\lambda_2|2))|\\
		&=|(p(\lambda_1|1)-p(\lambda_1|2))+(p(\lambda_2|1)-p_i(\lambda_2|2))|\\
		&=|p(\lambda_1|1)-p(\lambda_1|2)|+|p(\lambda_2|1)-p_i(\lambda_2|2)|
	\end{align*}
	holds due to the initial assumption $p(\lambda_1|1)\ge p(\lambda_1|2)$ and $p(\lambda_2|1)\ge p(\lambda_2|2)$, 
	we obtain
	\begin{align*}
		\sum_{\lambda'\in\Lambda'}|p(\lambda'|1)-p(\lambda'|2)|
		&=|p(\lambda'_1|1)-p(\lambda'_1|2)|+\sum_{\lambda'\in\Lambda'\backslash\{\lambda'_1\}}|p(\lambda'|1)-p(\lambda'|2)|	\\
		&=|p(\lambda_1|1)-p(\lambda_1|2)|+|p(\lambda_2|1)-p_i(\lambda_2|2)|+\sum_{\lambda\in\Lambda\backslash\{\lambda_1, \lambda_2\}}|p(\lambda|i)-p(\lambda|j)|\\
		&=\sum_{\lambda\in\Lambda}|p(\lambda|1)-p(\lambda|2)|.
	\end{align*}
	Now \eqref{eq:appB H} and \eqref{eq:appB M} prove
	\begin{align*}
		\frac{1}{4}\max_{\lambda\in\Lambda}\sum_ip(\lambda|i)+\frac{1}{8}\max_{i,j}\sum_{\lambda\in\Lambda}|p(\lambda|i)-p(\lambda|j)|
		&\le
		\frac{1}{4}\max_{\lambda'\in\Lambda'}\sum_ip(\lambda'|i)+\frac{1}{8}\max_{i,j}\sum_{\lambda\in\Lambda'}|p(\lambda'|i)-p(\lambda'|j)|\\
		&\le 1.
	\end{align*}

\bibliographystyle{quantum} 
\bibliography{ref_Bell_0228}
\end{document}